\documentclass[aps,prl,unsortedaddress,twocolumn,
]{revtex4-2} 

\usepackage{amsmath,amssymb}
\usepackage{graphicx}
\usepackage{dcolumn}
\usepackage{bm}
\usepackage[squaren,Gray,cdot]{SIunits}

\usepackage{mathtools} 

\newcommand{\Cn}{\tilde{\kappa}}
\newcommand{\DCn}{\Delta \tilde{\kappa}_c}
\newcommand{\Cc}{\kappa_c}
\newcommand{\Ccn}{\tilde{\kappa}_c}
\newcommand{\DC}{\Delta \kappa_c}

\usepackage{xcolor}
\usepackage{ulem}
\usepackage{colortbl}
\usepackage{color}
\usepackage[version=4]{mhchem}

\newcommand{\Bond}{B\hspace{-0.06em}o}

\DeclareMathOperator{\dmesure}{d}
\newcommand{\dx}{\dmesure\!}

\begin{document}

\preprint{APS}
\title{Equilibrium distribution of the liquid phase in an
  unsaturated granular material}
\author{Loredana Lazar, Cécile Clavaud,  and Axelle Amon}
\email{axelle.amon@univ-rennes.fr}
 \affiliation{Universit\'e Rennes, CNRS, IPR (Institut de Physique de Rennes) - UMR 6251, F-35000 Rennes, France}

\date{\today}
\begin{abstract}
In an unsaturated granular material, the spatial distribution of the liquid phase results from the competition between gravity and capillary forces. We show that, in the funicular regime, it can be described by a Boltzmann law, with static disorder playing the role of thermal agitation. We propose an approach based on a Langevin equation to derive this distribution, and compare our predictions with conductivity measurements giving the local water content as a function of height in a wet granular medium. We show that experimental data obtained with samples of different polydispersities collapse on a single master curve consistent with our model.
\end{abstract}

\maketitle
Wet granular materials are of large interest in industry (granulation in pharmaceutical processes, materials used for construction as concrete) as well as in soil sciences (stability of wet grounds, water contents and cycles in hydrology and agricultural-related studies). As such, they are the object of a renewed interest in the physics of granular materials community~\cite{mitarai2006,herminghaus2013}. 

These materials consist in a mixture of grains with a wetting liquid phase. When the porous matrix is not saturated, the distribution of liquid between the grains is governed by the competition between two effects: capillary forces that tend to localize the liquid at the level of the contacts between the beads and in the smallest pores, and gravity which drives liquid drainage. This competition can be quantified by a modified Bond number $\Bond_L = \frac{\rho g L}{\gamma / r}$, which is the ratio of the hydrostatic pressure over a characteristic length $L$ to the capillary pressure. Here, $\rho$ is the density of the liquid, $\gamma$ its surface tension, $g$ the standard acceleration of gravity, and $r$ the mean radius of the grains. At very low liquid content, in the \textit{pendular} regime, the liquid phase is located only at contacts or small gaps between the beads, and only forms capillary bridges. In this regime, gravity is not expected to play a role~\cite{mitarai2006,herminghaus2013}.  Indeed, in the pendular regime capillary bridges size is governed by $r$ so that the relevant choice for $L$ is $L \sim r$, giving the usual Bond number $\Bond_r = \left( r/\ell_c\right)^2$ with $\ell_c = \sqrt{\gamma/\rho g}$ the capillary length. Typically, for $r\simeq \unit{100}{\micro \meter}$ beads mixed with water, $\Bond_r \sim 10^{-3}$, meaning that capillary effects dominate the response. For larger amounts of liquid, the system enters the \textit{funicular} regime: some pores between the grains are filled, connecting capillary bridges to form clusters of liquid. Gravity is expected to play a role when these liquid clusters are large enough, and experiments performed in the funicular regime are generally done on samples of height $H$ small enough to be able to neglect gravity ($H/r \ll 1/\Bond_r$ so that $\Bond_H \ll 1$).

The relationship between the amount of liquid in the system and the pressure deficiency between the liquid and the gaseous phase has been the subject of numerous studies from the 1930's and the seminal works of Haines~\cite{haines1925,haines1927,haines1930} and Fisher~\cite{fisher1926}. There is an exact solution for the shape of a bridge of a given volume between two spheres~\cite{melrose1966,orr1975,willett2007} but not for more complex cluster shapes. Recently, an extensive study of liquid phase morphology in wet granular materials, based on X-ray tomography, provided a comprehensive picture of bridge formation, transition between the different regimes, and cluster morphology in the funicular regime, in systems small enough to neglect gravity effects~\cite{kohonen2004,scheel2008a,scheel2008b}. These studies confirm that for a wetting liquid, fluxes between bridges exist throughout the system, either because of the presence of a wetting film covering the whole solid phase~\cite{kohonen2004,lukyanov2012} or due to transfers through the gaseous phase~\cite{shahraeeni2012}. Consequently, in the absence of gravity and after a transient, all air/liquid interfaces equilibrate at the same curvature~$\kappa$. Moreover, they show that in the funicular regime, liquid clusters consist in capillary bridges fused together by the filling of the central pores between them. The pendular/funicular transition therefore occurs when the bridges connecting a grain to its neighbours are large enough to merge. Defining the liquid saturation $S$ as the ratio of the liquid volume to the pores one, the expected value of the critical saturation $S_c$ at this transition can be estimated, for monodisperse spheres, as:
\begin{equation}
    S_c=\frac{k}{2}\frac{v_{cb}^{(max)}}{\frac{4}{3}\pi r^3}\frac{\phi_G}{1-\phi_G}, \label{eq:Sc}
\end{equation}
where $k$ is the number of capillary bridges per particles, $\phi_G$ is the solid volume fraction and $v_{cb}^{(max)}$ is the maximum bridge volume for three capillary bridges connecting three identical beads in mutual contacts~\cite{herminghaus2013}. This bridge volume corresponds to a critical curvature $\kappa_c \simeq-4.46/r$ for vanishing contact angle~\cite{scheel2008a,herminghaus2013}. In the pendular regime, $S<S_c$ and increasing $S$ increases all the capillary bridges volume. In the funicular regime, $S>S_c$ and increasing $S$ results instead in the filling of pores between the bridges. Clusters of liquid in the funicular regime therefore have complex ramified structures with fractal characteristics. They are bounded by capillary bridges of curvature $\kappa_c$ independent of $S$~\cite{scheel2008a,badetti2018}, and their number and typical size increase with $S$~\cite{scheel2008a,herminghaus2013}. Importantly, due to disorder, a system can in fact adapt to a range of curvatures centered around $\kappa_c$, so that the allowed curvatures corresponding to bridges of maximal size span an interval of extension $2\Delta \kappa_c$. To give an order of magnitude based on the errorbar of~\cite{scheel2008a} fig.~3 ($r = \unit{140}{\micro \meter}$), $\gamma \DC/\rho g \sim \unit{10}{\centi \meter}$.

When working with large enough wet granular samples, for which $\Bond_H \gtrsim 1$ ($H/r \gtrsim 1/\Bond_r$), two natural and important questions therefore arise: 1) for what amount of liquid does gravity start playing a role?, and 2) what is the equilibrium spatial distribution of the liquid when it does? In this letter, we propose a physical model to answer these questions, and confront it to experimental measurements. In the first part of the following, we present our model, which yields a Boltzmann-type law for the vertical distribution of the excess saturation $\delta S = S-S_c$. We then present conductivity measurements giving the local degree of water saturation in a column of wet granular material. We show that all our experimental data collapse on a master curve based on our theoretical model, and discuss the role of the disorder in the system.

\begin{figure}[ht]
\centering
\includegraphics[width=\linewidth]{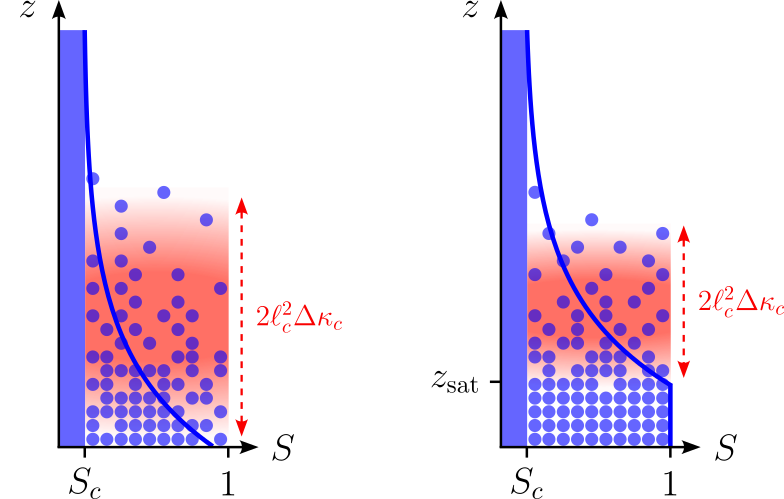}
\caption{Schematics of the distribution of the liquid phase with height. The liquid phase is represented as the sum of two contributions: a constant amount $S_c$ corresponds to the contribution of all the capillary bridges (the pendular skeleton). Additional liquid $\delta S$ is present as elements of liquid (blue circles) connecting the bridges to form clusters. At equilibrium, the curvature $\kappa$ of the air/liquid interfaces is directly linked to height $z$ and the funicular regime can exist only over a height $2\Delta z_c = 2 \ell_c^2 \DC$. Cases with (a) $S_i<S_\text{max}$ and (b) $S_i > S_\text{max}$ (see text for the definitions of $S_i$ and $S_\text{max}$).}
\label{fig:model}
\end{figure}

Let us consider a wet granular medium of height $H$, made of grains of radius $r \ll H$ mixed with a perfectly wetting liquid of density $\rho$ and surface tension $\gamma$. The system is taken to be in the funicular regime and to verify $\Bond_H \gtrsim 1$. The curvature of the air-liquid interfaces does not depend on whether they belong to a cluster or an isolated bridge~\cite{scheel2008a,herminghaus2013}. The liquid phase can therefore be decomposed into a \textit{pendular skeleton}, that is, a skeleton of capillary bridges all at their maximum volume, and the rest of the liquid, distributed as liquid elements filling some holes between liquid bridges, and considered as an excess amount with respect to the pendular skeleton. The liquid saturation $S$ can then also be decomposed into that of the pendular skeleton, which is $S_c$, and the excess saturation $\delta S = S - S_c$. We consider that the initial state of the system corresponds to a uniform distribution of the excess liquid elements, obtained for example by careful mixing, and that the initial local curvatures of the pendular skeleton are randomly drawn in a Gaussian distribution centered on $\kappa_c$ and of standard deviation $\Delta \kappa_c$. This initial state is out-of-equilibrium: because of the heterogeneous pressure field, transient flows will occur in the pendular skeleton in order to reach equilibrium in the gravity field, i.e. $\kappa(z) = (\Delta P_0 - \rho g z)/\gamma $ where $\kappa (z)$ is the curvature at equilibrium of the air/liquid interface at height $z$ and $\Delta P_0$ is the difference between the liquid pressure of the liquid at $z = 0$, at the bottom of the system, and the ambient air pressure. Note that large variations of $\kappa$ can be obtained with small variations of the bridges volume. The transient flow in the skeleton has two parts. First, its mean value corresponds to a downward flow because, at equilibrium, the $\kappa(z)$ gradient is associated to a bridge volume gradient. Second, random spatial fluctuations around this mean flow are expected due to the disorder of the initial pressure field. During this transient, the additional liquid elements flow together with the skeleton, exchanging liquid with it. At equilibrium, clusters can exist only at heights $z$ such that $\kappa (z) \in [\Cc - \DC; \Cc + \DC]$. Consequently, it is not possible to create a granular material in a funicular state of arbitrary size $H$ in a gravity field: the funicular regime can exist only over a height $\simeq 2\Delta z_c = 2 \ell_c^2 \DC$, and if $H \gg 2\Delta z_c$, the upper part of the system will necessary be in a pendular state (see fig.~\ref{fig:model}a). In the case when $\Delta z_c$ is too small to accommodate the excess of water, a saturated zone of height $z_\text{sat}$ will exist at the bottom of the system (see fig.~\ref{fig:model}b and SM). 

In the following, we neglect any permanent transfer between the skeleton and the elements (i.e. $S_c$ is independent of time), as well as the gradient of the volume of the skeleton bridges ($S_c$ is independent of space), and consider that the liquid saturation can be entirely described by the distribution of a finite number of liquid elements: the excess liquid elements. We consider that these elements are dragged by the flow and that their resulting movement is analogous to that of particles submitted to the superposition of a mean force and white noise, as described by a Langevin equation~\cite{huang}. The important specificity of our system is that the dynamics is transitory, and that both the downward flow and its fluctuations vanish together at equilibrium. We propose that this phenomenology can be described by an overdamped Langevin equation in which both the external force and the diffusion coefficient decrease with time and at the same rate $1/\tau$. The vertical position $z$ of each excess liquid element (of volume $v$) is therefore described by the following equation:
\begin{equation}
\Gamma \frac{\dx z}{\dx t} = -\frac{\partial V(z,t)}{\partial z} + \xi (t)
\end{equation}
with $\Gamma$ a friction coefficient, $V(z,t) = \rho v g z e^{-t/\tau}$ the time dependent potential, and $\xi(t)$ a random function modelling the fluctuations~\cite{huang}. This function satisfies $\langle \xi (t) \rangle = 0$ and $\langle \xi (t_1) \xi (t_2) \rangle = 2 D [(t_1 + t_2)/2] \Gamma^2 \delta (t_1 - t_2)$, where $\langle \cdot \rangle$ corresponds to the ensemble average on different initial realisations of both the disorder curvature field and the initial positions of the excess liquid elements. The diffusion coefficient $D(t) = D_0 e^{-t/\tau}$, as proposed above, is written as a time-dependent decaying function. The damping timescale $\tau$ is supposed to be very large compared to the typical fluctuation time scale $\tau_\text{fluc}$. Indeed, the ratio of these times can be estimated by the ratio of the mean flow rate, due to a pressure gradient $\rho g$, to its fluctuating part, driven by a pressure gradient of order $\gamma \DC/r$. This gives $\tau/\tau_\text{fluc} \sim \ell_c^2 \Delta \kappa_c/r \sim \Delta z_c/r$. In most situations, $\Delta z_c/r \gg 1$ and thus $\tau \gg \tau_\text{fluc}$. As long as this condition is fulfilled, the stochastic dynamics is well decoupled from the global slowing down of the process, and we can derive a Fokker-Planck equation for the probability $\mathcal{P}$ of an excess liquid element to be at height $z$ at time $t$~\cite{huang}:
\begin{equation}
    \frac{\partial \mathcal{P}}{\partial t} = - \frac{\partial}{\partial z} \left( \frac{\rho v g e^{-t/\tau}}{\Gamma} \mathcal{P} +  D_0 e^{-t/\tau} \frac{\partial \mathcal{P}}{\partial z}\right).
\end{equation}
At times $t \gg \tau$, $\mathcal{P}$ converges to the stationary solution $\mathcal{P}(z) = \mathcal{P}(0) e^{-\rho v gz/\Gamma D_0}$. For systems maintained at a constant temperature, $\Gamma D_0$ is the thermal energy of the bath. In our system, the evaluation of this term in not straightforward. The source of the fluctuations is the random initial pressure field. The strength of those fluctuations should govern the dispersion of the liquid elements. We thus expect $\frac{\Gamma D_0}{v} \sim \gamma \DC$ (see also SM). Finally, as $\delta S \propto \mathcal{P}(z)$, we predict the following form for the equilibrium distribution:
\begin{equation}
  \delta S(z) =  S(z) - S_c=\delta S_0 e^{- z / \Delta z_c} = \delta S_0 e^{- \Cn / \DCn}\label{eq:model}
\end{equation}
where $\Cn = rz/\ell_c^2$ and $\DCn = r \DC$. The $\delta S_0$ term depends on the total amount of liquid imposed in the experiment, that is, on the mean degree of saturation $S_i$ imposed at preparation. As $S(z) \leq 1$, there is a maximal saturation $S_\text{max} = S_c + \frac{1-S_c}{H}\int_0^H e^{-\frac{z}{\Delta z_c}} \dx z$, that can be accommodated by this exponential distribution. When $S_i>S_\text{max}$, an amount of liquid corresponding to $S_i - S_\text{max}$ will fill the bottom of the system, leading to the presence of a saturated zone of height $z_\text{sat}$ (fig.~\ref{fig:model}b).

To test our model, we perform experimental measurements of the local water content in a wet granular material, using a device initially designed to measure the volumetric fraction of the liquid phase in foams from conductivity measurements~\cite{feitosa2005}. The experimental device is a Plexiglas rectangular column $\unit{35}{\centi \meter}$ high and of section $\unit{2}{\centi \meter} \times \unit{2}{\centi \meter}$. Ten pairs of facing electrodes  are inserted regularly along the column's height. The conductance of the wet granular material is measured at the level of each pair of electrodes using  an impedance meter (Stanford Research Systems SR715). The voltage and frequency are the same as in~\cite{feitosa2005} ($\unit{1}{\volt}$ and $\unit{1}{\kilo \hertz}$) so that capacitive contributions are negligible and the measure gives the resistance of the material between each pair of electrodes.

The experiments are conducted using two sets of glass beads of respective radii $\unit{125 \pm 25}{\micro \meter}$ and $\unit{250 \pm 50}{\micro \meter}$, washed carefully and dried in an oven between each experiments. A fixed weight of grains is mixed with different amounts of a saline solution (distilled water + $\unit{0.4}{\mole \per \liter}$ $\ce{Na}\ce{Cl}$) in order to obtain wet granular materials with different mean saturations $S_i$ ranging from $2~\%$ to $58~\%$. The samples are then placed in the column and packed to obtain a reproducible solid volume fraction $\phi_G = \unit{0.53 \pm 0.01}{}$ for $r = \unit{125}{\micro \meter}$ and $\unit{0.56 \pm 0.01}{}$ for $r = \unit{250}{\micro \meter}$. Finally, the column is sealed to avoid evaporation effects. For each experiment, we wait  $\sim \unit{1}{\hour}$ for all the transitory flows to end. The resistance at the level of each pair of electrodes is then measured (average value over $\unit{1}{\hour}$).

The water saturation at height $z$ is deduced from the conductivity measurements using Archie's empirical law~\cite{archie1942}: $S = \sqrt{R_\text{sat}/R(S)}$, where $R_\text{sat}$ is the resistance of the saturated porous material and $R(S)$ its resistance for a saturation $S$. This empirical law holds at large enough values of $S$~\cite{han2009}, but there is a deviation at low saturations. This is to be expected, since the conduction paths must be of a different nature between the funicular and the pendular regime. We have checked that the mean saturation value over the height of the sample, as deduced from the electrical measurements, is equal to $S_i$ (see SM).

\begin{figure}[ht]
\centering
\includegraphics[width=\linewidth]{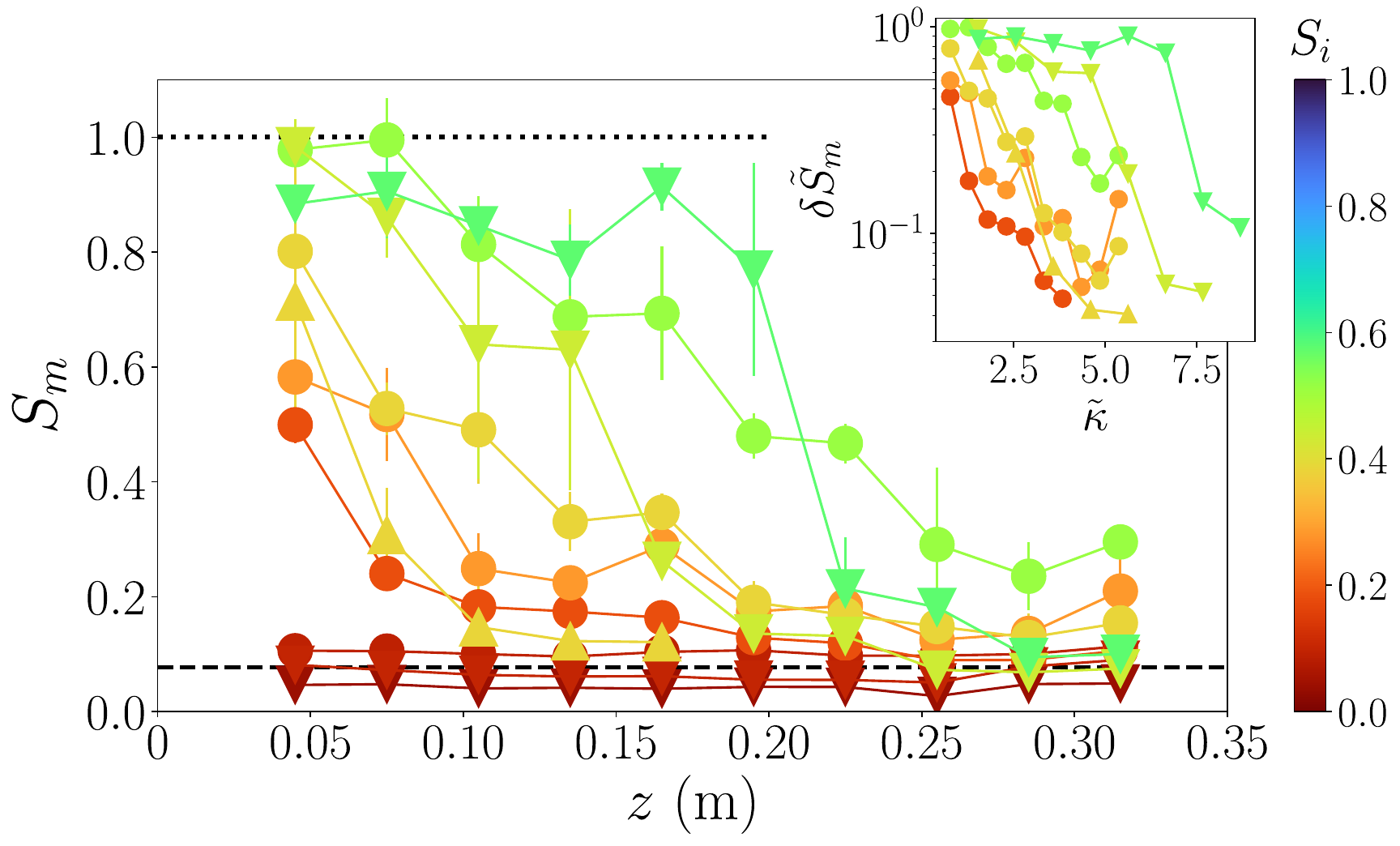}
\caption{$S_m(z) = \sqrt{\frac{R_\text{sat}}{R(z)}}$ computed from the resistivity measurements as a function of the electrodes positions $z$ for two bead sizes (circles: $r=125~\mu$m, $H=35$~cm; triangles: $r=250~\mu$m, $\blacktriangledown$ $H=35$~cm, $\blacktriangle$ $H=17.5$~cm). Errorbars: standard deviation on 3 different experiments. Color indicates the value of the liquid fraction $S_i$ imposed during the preparation (colorbar on the right). Dashed line: $S_c$ deduced from~\eqref{eq:Sc}. Insert: $\delta \tilde{S}_m = \frac{S_m - S_c}{1 - S_c}$ as a function of the non-dimensionalized curvature: $\Cn = r z/l_c^2$ in a semi-logarithmic~plot.}
\label{fig:saturation_vs_position}
\end{figure}

Figure~\ref{fig:saturation_vs_position} shows $S_m(z) = \sqrt{\frac{R_\text{sat}}{R(z)}}$, the measured saturation at height $z$, as a function of the position $z$ of the electrodes. We see that for $S_i \lesssim 10~\%$ and for both grain sizes, $S_m$ stays roughly constant. For $15~\% \lesssim S_i \lesssim 45~\%$, $S_m$ decreases continuously with $z$. For $S_i \gtrsim 50~\%$, $S_m$ first exhibits a plateau of value $\sim 1$ before decreasing. For the largest bead size, all the decreasing curves converge towards a common value $\simeq 0.1$. For the smallest beads, the limit values for the highest electrode are more dispersed, probably because transient flows in smaller pores are slower and the transient has not totally elapsed. For similar values of $S_i$, curves corresponding to the largest beads decrease faster with $z$ than the ones corresponding to the smallest beads.

In our model, we expect to observe an inhomogeneous distribution of the liquid phase for $S_i > S_c$. Our experimental observations point toward $S_c \simeq 0.1$. This value is significantly larger than that of $S_c = 0.047$ (resp. $S_c = 0.053$) predicted by~\cite{scheel2008a} using eq.~\eqref{eq:Sc} with $k=6$ and $\phi_G = 0.53$ (resp. $\phi_G = 0.56$). Such a value of $k$ corresponds to the number of direct contacts per beads at random close packing. Estimations from experimental~\cite{clark1967} and numerical studies~\cite{richefeu2006,badetti2018} point towards $k~\simeq 9-10$, due to the contribution of bridges between beads at close but not exact contact. Using $k=9.5$ in eq.~\eqref{eq:Sc}, we obtain $S_c \simeq 0.074$ (resp. $S_c \simeq 0.084$) for $\phi_G = 0.53$ (resp. $\phi_G = 0.56$). Using these values of $S_c$ we plot in insert $\delta \tilde{S}_m = (S_m - S_c)/(1 - S_c)$ as a function of $\Cn$ in a semi-logarithmic scale, showing that the measured $\delta S_m$ decreases exponentially. For $S_i \gtrsim 50~\%$, the plateau at small $z$ corresponds to saturation of the bottom of the column. The fact that $S_m(z) < 1$ in the saturated zone can be explained by the trapping of bubbles. The mean value of $S_c \simeq 0.08$ is represented as a dashed line in fig.~\ref{fig:saturation_vs_position}.

\begin{figure}[ht]
\centering
\includegraphics[width=\linewidth]{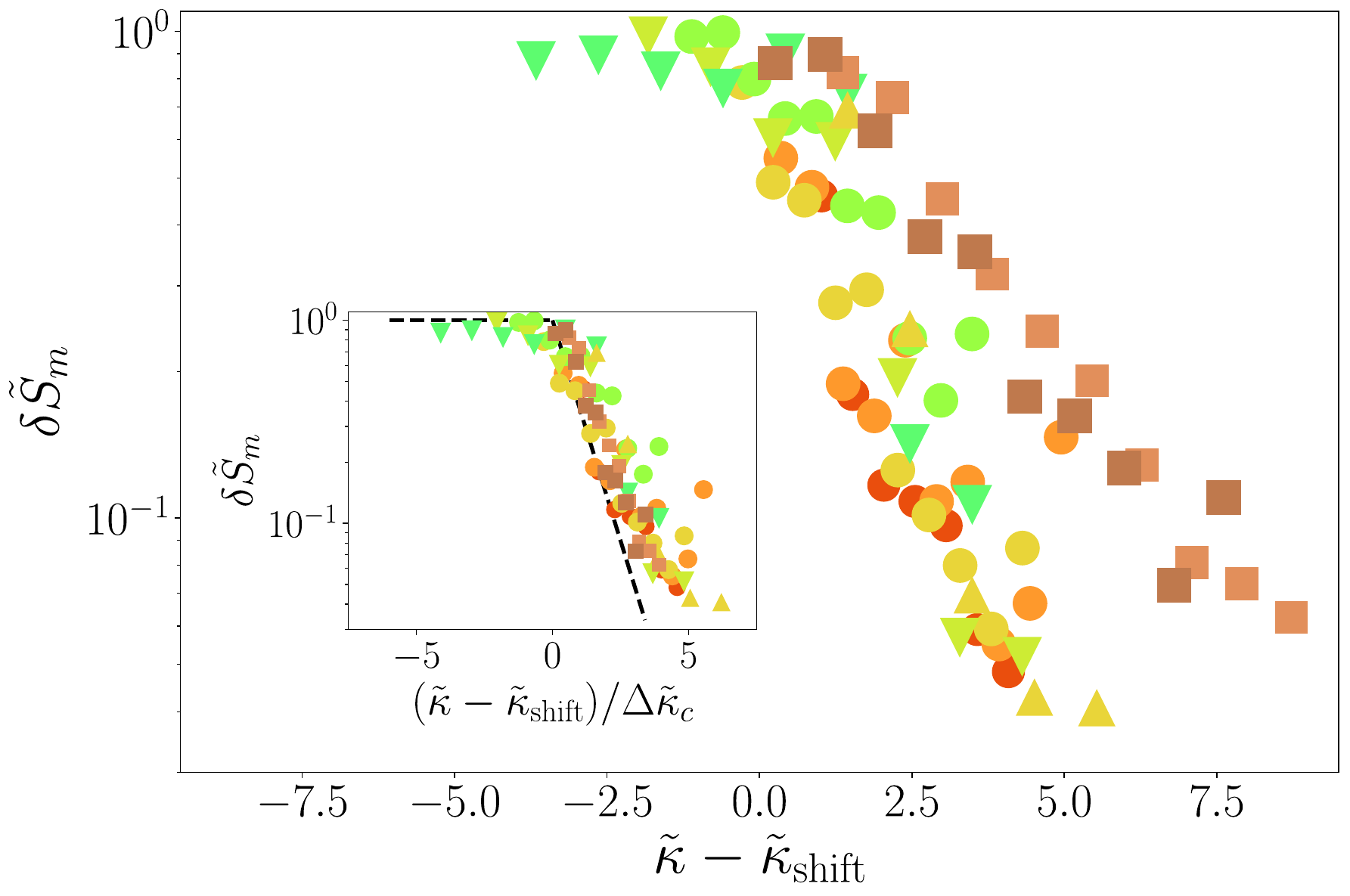}
\caption{Normalized excess of saturation $\delta \tilde{S}_m$ as a function of $\Cn - \Cn_\text{shift}$ for three bead sizes (circles: $r=125~\mu$m, $H=35$~cm; triangles: $r=250~\mu$m, $\blacktriangledown$ $H=35$~cm, $\blacktriangle$ $H=17.5$~cm; squares: $r=200~\mu$m, $H=35$~cm). The color scale for the preparation saturation degree $S_i$ is the same as the one of fig.~\ref{fig:saturation_vs_position} for the $125~\mu$m and $250~\mu$m beads but is different for the $200~\mu$m ones to clearly distinguish the data. Insert: $\delta \tilde{S}_m$ as a function of $(\Cn - \Cn_\text{shift})/(|\Ccn| \frac{\Delta r}{r})$.}
\label{fig:shifted_curves}
\end{figure}

The main parameter of our model is the energy per unit volume $\frac{\Gamma D_0}{v}$ that we have supposed to be equal to $\gamma \DC$. To further test the robustness of eq.~\eqref{eq:model}, we want to estimate this parameter from the properties of our porous material. The width of the range of authorized curvatures has different origins~\cite{mani2015}: packing disorder, polydispersity, contact angle hysteresis\dots We suppose that in our system, polydispersity is the governing factor for this dispersion, and that $\left| \frac{\DC}{\Cc}\right| \simeq \frac{\Delta r}{r}$. Recalling that $\Ccn = -4.46$ and knowing that $\frac{\Delta r}{r} \simeq 0.2$ for the two sets of beads, we compute $\DCn \simeq 0.9$, a value close to the slope of the curves in the insert of fig.~\ref{fig:saturation_vs_position}. From this, we compute both the value of $\delta S_0$ when $S_i<S_\text{max}$ and that of $z_\text{sat}$ when $S_i>S_\text{max}$ (detailed calculation in SM). Fig.~\ref{fig:shifted_curves} shows $\delta \tilde{S}_m$ as a function of $\Cn - \Cn_\text{shift}$, with $\Cn_\text{shift} = \DCn \ln \delta \tilde{S}_0$ (respectively $r z_\text{sat}/\ell_c^2$) when $S_i < S_\text{max}$ (resp. $S_i >S_\text{max}$). Points and triangles correspond to the data in the insert of fig.~\ref{fig:saturation_vs_position}, with the same color code. The brown squares correspond to measurements done with another set of beads of higher polydispersity ($r \pm \Delta r = \unit{200\pm 100}{\micro \meter}$, therefore $\DCn \simeq 2.25$). Note that while the computation of $\Cn_\text{shift}$ when no part of the column is saturated is exact, the value of $z_\text{sat}$, the height of the saturated area for $S_i > S_\text{max}$, is only an estimation.

The sets of grains used in our main experiments (circles, $\unit{125\pm 25}{\micro \meter}$, and triangle, $\unit{250 \pm 50}{\micro \meter}$) have the same relative size dispersion $\frac{\Delta r}{r} \simeq 0.2$. Interestingly these points approximately collapse on a common curve in fig.~\ref{fig:shifted_curves}. The data corresponding to the granular material of higher polydispersity ($\unit{200\pm 100}{\micro \meter}$, $\frac{\Delta r}{r} \simeq 0.5$) shows a slower decrease than the less polydisperse cases. Insert of fig.~\ref{fig:shifted_curves} shows the experimental data from all three sets of grains plotted as a function of $(\Cn - \Cn_\text{shift})/(|\Ccn| \frac{\Delta r}{r})$, where the dashed line corresponds to the model without any adjustable parameter. We see that all the data nicely collapse on a common master curve. The difference in slope between the experimental data and the model is consistent with an underestimation of disorder. Such a discrepancy is expected as other sources of disorder should be involved in addition to polydispersity.

In conclusion, we have shown that when a granular material is prepared by homogeneously mixing a wetting liquid with glass beads at a saturation degree $S_i > S_c \simeq 0.1$, partial drainage occurs in the porous material, leading to a final Boltzmann-type distribution of the liquid content. This distribution describes the excess of liquid $\delta S = S - S_c$ that cannot be accommodated by an increase of the mean capillary bridges size, and thus leads to the formation of liquid clusters distributed in the system. The width of this distribution (i.e. the analog of the thermal energy) is the dispersion of the possible curvatures, which is governed by the disorder of the granular packing. This disorder exists because the preparation of a wet granular material implies a careful mixing, setting as an initial condition a distribution of the local Laplace pressures. Because of the existence of wetting films connecting the entire fluid phase, gravity always plays a role, in the sense that it sets the values of the curvatures in the steady state everywhere in the system. While the subsequent gradient of bridges volume is small (and within the errorbars of our measurements) in the pendular state, an heterogeneity of the steady distribution of water is observed in the funicular sate. Interestingly, our approach gives a physical argument to the phenomenological laws proposed in hydrology and soil mechanics to fit the curves linking the pressure deficiency to the water content (the so-called \textit{soil-water characteristic curves}, SWCC)~\cite{fredlund1993,dingman1994}. Indeed, the typical approach in soil science studies consists in fitting $S-S_r$ where $S_r$ is called the residual degree of saturation and is a fitting parameter. Most of the SWCC curves display an exponential decrease far from the saturated zone~\cite{tani1982,fredlund1993,dingman1994}. Our model gives a physical origin to some of the parameters used in those phenomenological models. Although it will certainly present some limits at very high polydispersity (for example when clay is present in the material), it provides a useful tool to predict the liquid distribution in experiments based on wet glass beads.\\


The authors thank Isabelle Cantat, Jérôme Crassous and Arnaud Saint-Jalmes for inspiring discussions. We also thank A. Saint-Jalmes for loaning the experimental device, Patrick Chasle and Florian Scholkopf for technical support and Lény Cancho-Garry for additional measurements. This work has been partly supported by the French National Research Agency through the project ModAFroSt (Grant ANR-24-CE30-4668-01). L.~L. acknowledges support from Bretagne Région (ARED program).

\bibliography{biblio_drainage}
\bibliographystyle{apsrev4-2}

\end{document}